\documentclass[letterpaper, 10 pt, conference]{ieeeconf}  

\usepackage{graphicx}
\usepackage{authblk}
\usepackage{cite}

\IEEEoverridecommandlockouts                              
\title{\LARGE \bf FELIX based readout of the Single-Phase ProtoDUNE detector}

\author{
Andrea Borga$^{1}$,
Eric Church$^{2}$,
Frank Filthaut$^{3}$,
Enrico Gamberini$^{4}$,
Paul de Jong$^{5}$,
Giovanna Lehmann Miotto$^{4}$,\\
Frans Schreuder$^{1}$,
J\"orn Schumacher$^{4}$,
Roland Sipos$^{4}$,
Milo Vermeulen$^{1}$,
Kevin Wierman$^{2}$,
Lynn Wood$^{2}$
\thanks{$^{1}$Andrea Borga, Frans Schreuder and Milo Vermeulen are with Nikhef, Amsterdam, The Netherlands}
\thanks{$^{2}$Eric Church, Kevin Wierman and Lynn Wood are with Pacific Northwest National Laboratory, Richland, Washington, USA}%
\thanks{$^{3}$Frank Filthaut is with Radboud University, Nijmegen and Nikhef, Amsterdam, The Netherlands}%
\thanks{$^{4}$Enrico Gamberini, Giovanna Lehmann Miotto, J\"orn Schumacher and Roland Sipos are with CERN, Geneva, Switzerland}%
\thanks{$^{5}$Paul de Jong is with Nikhef and the University of Amsterdam, Amsterdam, The Netherlands}%
}

\begin{document}

\maketitle
\thispagestyle{empty}
\pagestyle{empty}

\begin{abstract}
Large liquid argon Time Projection Chambers have been adopted for the DUNE experiment's far detector, which will be composed of four 17\,kton detectors situated 1.5\,km underground at the Sanford Underground Research Facility. This represents a large increase in scale compared to existing experiments. Both single- and dual-phase technologies will be validated at CERN, in cryostats capable of accommodating full-size detector modules, and exposed to low-energy charged particle beams. This programme, called ProtoDUNE, also allows for extensive tests of data acquisition strategies.
The Front-End LInk eXchange (FELIX) readout system was initially developed within the ATLAS collaboration and is based on custom FPGA-based PCIe I/O cards, connected through point-to-point links to the detector front-end and hosted in commodity servers. FELIX will be used in the single-phase ProtoDUNE setup to read the data coming from 2560 anode wires organized in a single Anode Plane Assembly structure. With a continuous readout at a sampling rate of 2 MHz, the system must deal with an input rate of 96\,Gb/s. An external trigger will preselect time windows of 5\,ms with  interesting activity expected inside the detector. Event building will occur for triggered events, at a target rate of 25\,Hz; the readout system will form fragments from the data samples matching the time window, carry out lossless compression, and forward the data to event building nodes over 10\,Gb/s Ethernet.
This paper discusses the design and implementation of this readout system as well as first operational experience.
\end{abstract}

\section{INTRODUCTION}
ProtoDUNE-SP\cite{Abi:2017aow} is the single-phase DUNE Far Detector prototype that is under construction and will be operated at the CERN Neutrino Platform (NP) starting in 2018. ProtoDUNE-SP represents a crucial part of the DUNE effort towards the construction of the first DUNE 10 kton fiducial (17\,kton total) liquid argon (LAr) mass Far Detector module. With a total LAr mass of 0.77 kton, it represents the largest monolithic single-phase LAr Time Projection Chamber (TPC)  built to date. It is housed in an extension to the EHN1 hall in the North Area, where the CERN NP is providing a new dedicated charged-particle test beamline. ProtoDUNE-SP aims to take its first beam data in the second half of 2018.

\begin{figure}[thpb]
\centering
\includegraphics[width=.4\textwidth]{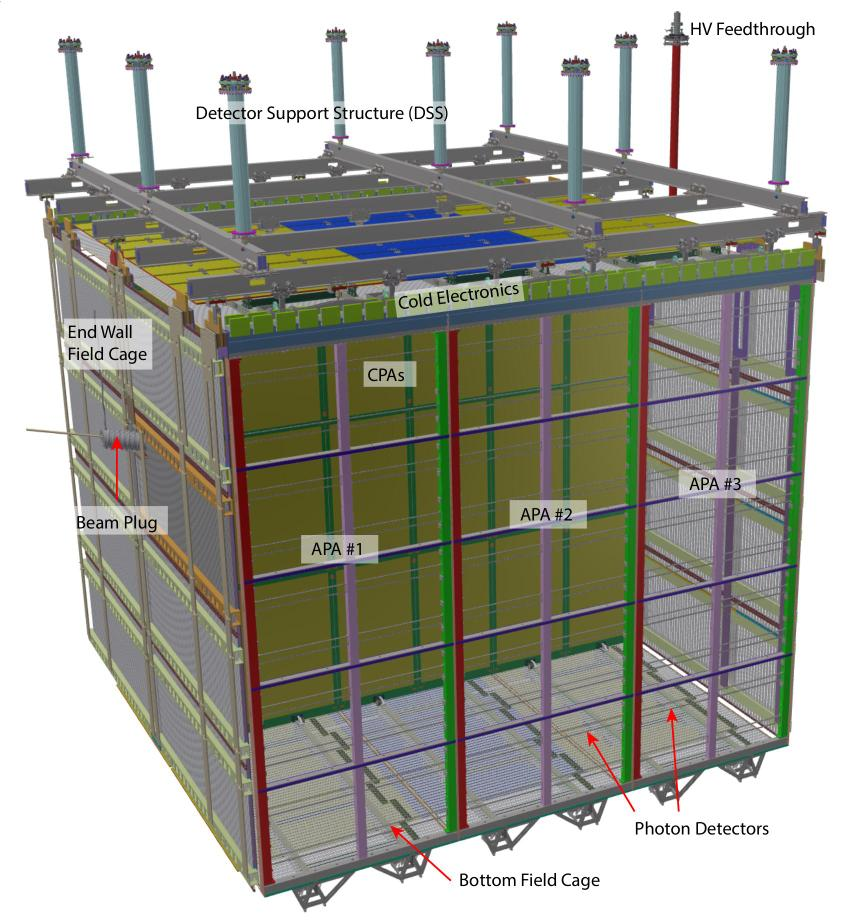} 
\caption{The major components of the ProtoDUNE-SP TPC.}
\label{PDSP_TPC}
\end{figure}

The ProtoDUNE-SP TPC, illustrated in Figure \ref{PDSP_TPC}, comprises two drift volumes, defined by a central cathode plane that is flanked by two anode planes, each at a distance of 3.6\,m, and a field cage (FC) that surrounds the entire active volume. The active volume is 6\,m high, 7\,m wide and 7.2\,m deep (along the drift direction). Each anode plane is constructed of three adjacent Anode Plane Assemblies (APAs) that are each 6\,m high by 2.3\,m wide in the installed position. Each APA consists of a frame that holds three parallel planes of induction and collection wires on each of its two faces for a total of 2560 channels; the wires of each plane are oriented at different angles with respect to those making up the other planes of the same face, to enable 3D reconstruction.

The readout of the TPC wires, prior to being received by the DAQ system, consists of cold electronics (CE) mounted on the APAs inside the cryostat and the warm electronics outside the cryostat on the flange. CE data are received on the Warm Interface Boards (WIBs) which are situated on the top of the flanges. Each WIB multiplexes the data to four 4.8\,Gb/s (or two 9.6\,Gb/s) lines that are sent over optical fibers to the DAQ. Two systems are used to receive data from the WIBs. The baseline system is based on Reconfigurable Computing Elements (RCE)~\cite{Herbst:2016prn} which are used to read out 5 of 6 APAs, while the alternative system described here is based on the Front-End Link EXchange (FELIX)~\cite{Anderson:2016lfn} technology and is used to receive the data from the remaining APA.

\section{FELIX-BASED READOUT}
The main driver of the FELIX concept is the firm belief that a thin interface managing the interaction with detector front-end links and injecting data into commodity servers at an early stage of the DAQ chain provides the flexibility that is required for the optimization and maintenance of long term and long lifetime systems. The FELIX I/O card provides a simple point-to-point interface to the detector front-end, supporting a 8b/10b encoded serial protocol at 9.6\,Gb/s. Using the PCIe format allows all data to be transferred to the host memory. This solution leverages the fast evolution of multi-core server performance, the possibility of using the large available host memory and the optimal choice of high performance networking for data dispatching.

\subsection{Topology of the readout system}
The FELIX I/O card interfaces with its host PC through 16-lane PCIe Gen3. It transfers the incoming WIB data directly into the host PC's memory using a continuous DMA transfer accomplished through the Wupper~\cite{Wupper} engine. The host PC runs a software process, called felixcore~\cite{felixcore}, that publishes selected data to any client subscribing to it, based on logical link identifiers. In ProtoDUNE the clients to the felixcore application are the BoardReader processes, which are part of the artdaq~\cite{Biery:2013cda} framework used for the ProtoDUNE DAQ dataflow system.

From a hardware point of view the FELIX and BoardReader hosts in use are based on a dual socket Intel\textsuperscript{\textcopyright} Xeon\textsuperscript{\textcopyright} Processors (E5-2620 v4 2.1 GHz), equipped with a Mellanox Technologies MT28800 Family [ConnectX\textsuperscript{\textcopyright}-5 Ex] 2x100 Gb/s NIC. A single FELIX I/O card receives data from a whole APA over 10 links and is hosted in one server. On the other hand, final performance benchmarking is still needed to establish whether a single host for the BoardReader processes receiving data from the FELIX will be sufficient, or a second host will be required. The output of selected data towards the DAQ event builder is carried out over 10 Gb/s Ethernet.

\begin{figure}[thpb]
\centering
\includegraphics[width=0.8\linewidth]{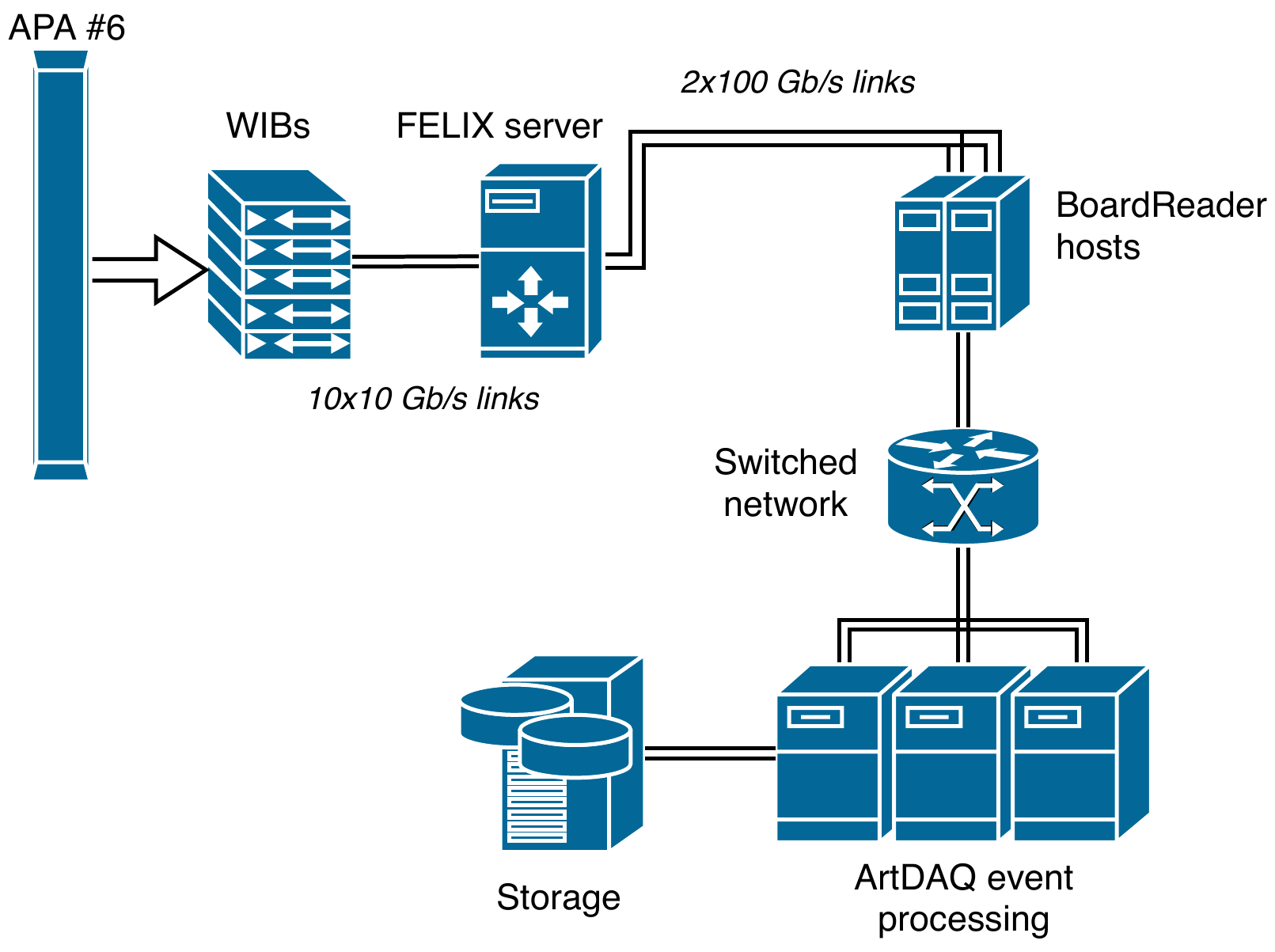}
\caption{Overview of the FELIX data acquisition chain for ProtoDUNE-SP.}
\label{felixDaqOverview}
\end{figure}
\subsection{WIB data volume and structure}
\label{sec:wib-data}
In the ProtoDUNE-SP context, the readout system must be able to cope with the bandwidth of data transfers from the WIBs. The WIBs will send data to FELIX at a 2\,MHz frame rate per optical link. Each frame contains 120 32-bit words, leading to a payload rate of 7.68\,Gb/s; the additional overhead from 8b/10b encoding leads to a total transfer rate of 9.6\,Gb/s. Each link represents 256 channels; FELIX therefore needs to read from 10 input links, corresponding to a total payload rate of 76.8\,Gb/s. 

The readout system must buffer the incoming data until a data request is received from the event building farm, and then transfer the data contained in a 5\,ms time window, centered around the trigger timestamp. The DAQ system is designed for a target trigger rate of 25\,Hz. An overview of the described readout system is seen in Figure \ref{felixDaqOverview}.

In the case of FELIX, the WIBs combine data from two Front-End Mother Boards (FEMBs) into one frame. Since each FEMB sends out 128 channels at a rate of 2\,MHz, FELIX frames contain 256 channel values in total, divided over four blocks. The channel values each take up 12 bits, but are cut up and rearranged to be byte-aligned. FEMBs may individually fail, but the WIB will continue sending fixed size fragments towards the FELIX.

Apart from the ADC values, the WIB frames contain a WIB header as well as four COLDATA headers. As the name suggests, the information in the WIB header is added by the WIB. It contains identifier data such as the APA number, WIB number and WIB output fiber number that the data originated from. Together, these uniquely identify the origin of the data encompassed by the frame. Additionally, there are several error fields available for the WIB to pass error flags along. Lastly, the WIB header contains a 63-bit timestamp, which is generated from a detector wide timing system and increments each 20 ns. From one frame to the next, this timestamp is expected to increment by 25 such intervals, given the 2\,MHz frame rate. 

The channel values in each frame are divided into four COLDATA blocks, which each contain 64 values and have a COLDATA header each. This header
contains additional error fields as well as a COLDATA convert count, which is expected to increment between consecutive frames from the same source. These counters are only supposed to be identical when coming from the same FEMB and can therefore differ within a frame, which contains data from two separate FEMBs. Lastly, the COLDATA headers contain its own checksums.

Each frame contains a CRC-20 checksum generated by the WIB. It is used by FELIX to verify the frame's data integrity and is then discarded before the frame is passed to the FELIX host PC.

\subsection{FELIX firmware}
The FELIX firmware design has been described in Ref.~\cite{felixfw}. In order to be able to sustain the high rate of incoming fragments (2 MHz) and the high throughput requirements, the firmware is being modified in several parts, specifically for ProtoDUNE:
\begin{itemize}
\item The fixed frame size allows the (fixed) block size used for DMA transfers to be matched to the frame size or a multiple thereof, a multiple of 6 being preferred at present as it allows to avoid parsing each individual time slice in the software.
\item Data from each individual link are DMA'd into different memory areas on the host.
\end{itemize}

These three changes allow to simplify considerably the felixcore software running on the FELIX host. 

\begin{figure}[thpb]
\centering
\includegraphics[width=.48\textwidth]{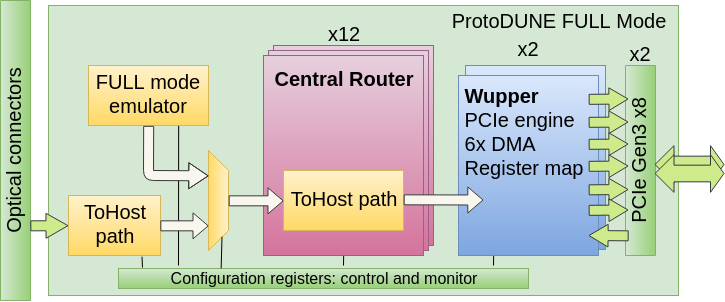}
\caption{Main firmware logical blocks for 8b/10b data mode}
\label{FELIX_FW}
\end{figure}

\subsection{The felixcore application}
The felixcore application is in charge of routing data from the detector to networked software clients and vice versa~\cite{felixcore}. It has been designed with the aim of being  generic and unaware of the data that it routes. It supports several networking back-ends, such as zeromq, TCP/IP and Infiniband verbs. 

In ProtoDUNE only uni-directional traffic is used (from detector) and the data fragments have a fixed size. These two features have been used in order to optimize the performance of the felixcore application, thus achieving the required data throughput. In particular the routing of data was simplified by dedicating a thread to every physical link and by introducing scatter-gather techniques in order to be able to send large messages over the network without performing any data copies.

\section{DAQ SOFTWARE LAYER}

The ProtoDUNE use-case has challenging requirements for the software downstream of the FELIX readout, since it has to support the full data rate, perform trigger matching and lossless data compression. Previous experiments such as MicroBooNE~\cite{Acciarri:2017sde} have suggested that for a sufficiently low electronics noise level, a compression factor of 4 is feasible for the ProtoDUNE-SP data. Nevertheless, general-purpose compression algorithms are less efficient, especially if data are not re-organized prior to compression.

\subsection{BoardReader implementation}
The BoardReader implementation for the FELIX based readout integrates the NetIO~\cite{Schumacher:2017gvf} messaging layer that subscribes to the felixcore application into the artdaq~\cite{Biery:2013cda} framework, which is used as ProtoDUNE's data acquisition software framework. In particular, each BoardReader subscribes to data from one WIB link.
Its main tasks are to :
\begin{itemize}
\item store all incoming data into a circular buffer;
\item receive trigger information (timestamp and event identifier);
\item form a DAQ fragment by matching the data in the circular buffer and the timestamp of the trigger (5\,ms of data around the trigger);
\item compress the content of the DAQ fragment (as discussed in more detail below);
\item and pass on the DAQ fragment to the event builder.
\end{itemize}

In order to achieve the required performance, particular care has been put into the implementation of the customized part of the BoardReader application, avoiding as much as possible dynamic memory allocation.
Internal elements are accessed through unique pointers and the access functionalities strictly avoid copies. Every link has dedicated subscriber threads that populate single producer single consumer (SPSC) queues, using the lock-free implementation from the Folly~\cite{folly} library. An SPSC queue can be used for communication with thread which services a hardware device (wait-free property is required), or when there are naturally only one producer and one consumer. As our solution has a single extractor (trigger-matcher) for each circular buffer, using a lock-free implementation is a straightforward choice. The thread flow and the utilization of the circular buffer are shown in Fig 4.
\begin{figure}[thpb]
\centering
\includegraphics[width=.48\textwidth]{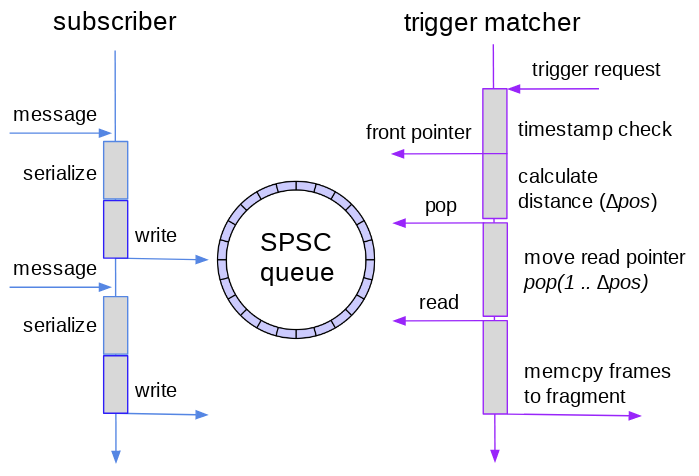} 
\caption{The logic of producer and consumer threads, and their utilization of the queue.}
\label{FELIX_SW}
\end{figure}

The solution relies heavily on the standard library of C++ (std11). Move semantic is used to avoid memory copies, and to directly stream the data from the network interface card's ring buffer to the queues. Multi-threading features ensure the proper synchronization of the trigger matching threads, and also to comply with the internal state machine of artdaq. The implementation also focuses on flexibility, as the topology of the queues and links is scalable by the BoardReader configuration.

\subsection{Compression}
Two constraints exist for lossless compression within the FELIX system. First, the compression factor must be effective.
The target compression factor for the ProtoDUNE-SP data was set to 4
and is incorporated in storage hardware projections.
Second, the compression algorithm should keep up with the trigger rate. At 25 Hz, each BoardReader process has only 40\,ms to handle its current batch of data.
Since the data compression stage is the computationally most demanding one, a fast solution is essential.

The speed of the compression procedure is optimized using dedicated Intel\textsuperscript{\textregistered} QuickAssist Technology (QAT)\cite{qat} hardware. 
This employs the DEFLATE algorithm by default, which consists of a sliding window compression and a Huffman compression stage.
In the former, repeated bit strings are replaced with references to a previous occurrence.
The latter then replaces frequently occurring bytes with shorter bit strings.
This allows a reduction of the time required for the compression of one fragment's data to approximately 4\,ms.
For unmodified, frame-by-frame input data, the achieved compression factor is well below the desired factor of 4.
This is due in part to the non-contiguous storage of ADC data for individual channels in subsequent frames, but more importantly to the fact that part of the ADC values are cut up in the frame data, as mentioned in Sec.~\ref{sec:wib-data}.
The BoardReader process will therefore feature a data re-ordering stage, where the ADC values are re-constructed in 16-bit words and re-ordered so that the ADC values for all 10,000 subsequent digitization time slices are contiguous in memory.
Using a simulated electronics noise of approximately 4 ADC counts~\cite{Acciarri:2017sde} on average, a compression factor of 3.7 is reached.

\section{Summary}

The ProtoDUNE-SP detector is a 770\,ton LAr detector intended to validate the single-phase LAr Time Projection Chamber technology at the full scale of the DUNE experiment, and expects to receive beam from the CERN SPS accelerator in the second half of 2018.
One of its six Anode Plane Assemblies, representing 2560 anode wires, will be read out using the FELIX system.

The FELIX readout system is based on the concept of having a thin interface between the front-end of a detector and commodity hardware.
The current FELIX I/O card receives 96\,Gb/s of data over 10 links and uses 16-lane PCIe Gen3 to copy it to the FELIX host PC's memory.

The input data rate can be sustained using a firmware and software modified from its original version used in the ATLAS experiment, with a separate DMA transfer for each input link's data, and using larger block size matched to the input frame size. This is combined with the use of scatter-gather techniques to send the data to BoardReader processes running on separate hosts.

These BoardReader processes must perform trigger matching and lossless data compression by a factor of 4. The requirements on the compression, which is the most computationally demanding step, have largely been met by re-formatting the data in software and subsequently carrying out a hardware accelerated compression. They are subsequently packaged in the artdaq fragment format and forwarded to the artdaq based event building framework.

\addtolength{\textheight}{-12cm}   




\section*{Acknowledgment}
We thank our colleagues in the data acquisition group of the single-phase ProtoDUNE setup, within which this project is embedded, for their support in integrating the FELIX based readout;
and the CERN openlab team, for providing support for our hardware accelerated compression studies.
We gratefully acknowledge the assistance and support from the FELIX developers team on the ATLAS experiment, without whom this project would not have been possible.

\bibliography{mybib}{}
\bibliographystyle{plain}

\end{document}